\documentclass[a4paper,11pt]{article}
\usepackage{pos}
\usepackage{amsfonts}
\usepackage{caption}
\usepackage{subcaption}
\usepackage{xcolor}
\usepackage{graphicx}

\usepackage{amsbsy}
\usepackage{amsmath}

\newcommand{\mb}[1]{\mathbf{#1}}

\newcommand{\qqquad}{\quad\qquad}
\newcommand{\half}{\frac{1}{2}}
\newcommand{\nn}{\nonumber}

\title{Random Matrix Theory for Stochastic Gradient Descent}
\author*[a]{Chanju Park}
\author*[a]{Matteo Favoni}
\author[b]{Biagio Lucini}
\author[a]{Gert Aarts}

\affiliation[a]{Department of Physics, Swansea University,\\
  Swansea, SA2 8PP, United Kingdom}

\affiliation[b]{Department of Mathematics, Swansea University (Bay Campus),\\
  Swansea, SA1 8EN, United Kingdom}

\emailAdd{chanju.b.park@gmail.com}
\emailAdd{matteo.favoni@swansea.ac.uk}
\emailAdd{b.lucini@swansea.ac.uk}
\emailAdd{g.aarts@swansea.ac.uk}

\abstract{
    Investigating the dynamics of learning in machine learning algorithms is of
    paramount importance for understanding how and why an approach may be
    successful. 
    The tools of physics and statistics provide a robust setting for such
    investigations. 
    Here we apply concepts from random matrix theory to describe stochastic
    weight matrix dynamics, using the framework of Dyson Brownian motion. 
    We derive the linear scaling rule between the learning rate (step size) and
    the batch size, and identify universal and non-universal aspects of weight
    matrix dynamics. 
    We test our findings in the (near-)solvable case of the Gaussian Restricted
    Boltzmann Machine and in a linear one-hidden-layer neural network.
    }

\FullConference{The 41st International Symposium on Lattice Field Theory (LATTICE2024)\\
 28 July - 3 August 2024\\
Liverpool, UK\\}

\begin{document}
\maketitle

\section{Introduction}

  Machine learning (ML) and artificial intelligence (AI) can provide powerful
  tools for the scientific community, as demonstrated by the recent Nobel Prize
  in Chemistry.
  Reversely, insights from traditional physics theories also contribute to a
  deeper understanding of the mechanism of learning. 
  Ref.~\cite{Carleo_2019} contains a broad overview of the successful
  cross-fertilisation between ML and the physical sciences, covering a number
  of domains.
  One way to mitigate against possible scepticism %in the scientific community
  with regard to using ML as a ``black box'' is by unveiling the dynamics of
  training (or learning) and explaining how the relevant information is engraved
  in the model during the training stage.

  To further develop this programme, we study here the dynamics of first-order
  stochastic gradient descent as applied to weight matrices, reporting and
  expanding on the work presented in Ref.~\cite{Aarts:2024wxi}.
  When training ML models, weight matrices are commonly updated by one of the
  variants of the stochastic gradient descent algorithm.
  The dynamics can then be decomposed into a drift and a fluctuating term, and
  such a system can be described by a discrete Langevin equation.
  The dynamics of stochastic matrix updates is richer than the dynamics for
  vector or scalar quantities, as captured by Dyson Brownian motion and random
  matrix theory (RMT), with the appearance of universal features for the
  eigenvalues \cite{Wigner-1, Wigner-2,Dyson-1, Dyson-2,
  Dyson-3,Dyson-4,Meh2004}. 
  Earlier descriptions of the statistical properties of weight matrices in
  terms of RMT can be found in e.g.\ Refs.~\cite{Martin-2019, Baskerville}, but
  here we specifically focus on RMT effects for stochastic gradient dynamics
  via Dyson Brownian motion, which leads to additional Coulomb-type repulsion
  between eigenvalues due to the Vandermonde determinant.
  Importantly, we have shown that in this framework a specific combination of
  hyperparameters of the optimiser, namely the ratio of the learning rate (or
  step size) and the batch size, naturally arises as a scaling factor
  determining the strength of the fluctuations in the process
  \cite{Aarts:2024wxi}.
  In fact, this specific combination had already been observed at an empirical
  level in practical ML training and dubbed the {\em linear scaling rule}
  \cite{Goyal-1, Smith-1, Smith-2, Smith-3}. We derived this relation from
  first-principle matrix dynamics \cite{Aarts:2024wxi}.

  In this contribution, we first summarise the results of
  Ref.~\cite{Aarts:2024wxi} and then present some new results for a simple
  linear neural network with one hidden layer.
  Related results for a nano-GPT can be found in Ref.~\cite{Aarts:2024qey}.

\section{Langevin equation for stochastic gradient descent}

  Stochastic gradient descent (SGD) is one of the most commonly used first-order
  gradient optimisation algorithms in the ML community.
  Given an objective function $\mathcal{L} (W)$ depending on a (collection of)
  weight matrices $W$, the optimal state that minimises the objective function
  is found by searching for the stationary point of the first-order equation,
  \begin{align}
    W_{n+1} = W_{n} - \alpha \left\langle \Delta_{p} \right\rangle_{p \in \mathcal{B}},
    \label{eq:SGD}
  \end{align}
  where
  \begin{align}
    \left\langle \Delta_p \right\rangle_{p \in \mathcal{B}}
    \equiv \frac{1}{|\mathcal{B}|} \sum_{p \in \mathcal{B}} \Delta_p,
    \qqquad \Delta_p \equiv \frac{\partial \mathcal{L}}{\partial W_n} \bigg|_p.
    \label{eq:batch_grad}
  \end{align}
  Here $\alpha$ is the learning rate and the gradient $\Delta_p$ of the objective
  function $\mathcal{L}(W)$,  depending on the current state $W_n$, is calculated
  for each data point $p$ and then averaged over the data points within a
  (mini-)batch $\mathcal{B}$.

  Unlike in standard gradient descent, the input data is split into a number of
  small mini-batches and stochasticity is introduced due to the effect of
  having finite sample sizes.
  Assuming that the input data is well standardised, each gradient within a
  mini-batch is an i.i.d.\ random variable as each data point in the mini-batch
  is randomly sampled from the total dataset.
  As the measured batch gradient is the average of i.i.d.\ variables, we can
  use the central limit theorem to write $\left\langle \Delta_p \right\rangle$
  in terms of the mean gradient of the batch and its fluctuation,
  \begin{align}
    \left\langle \Delta_p \right\rangle_{p \in \mathcal{B}}
    = \mathbb{E}_{\mathcal{B}}[\Delta_p] + \frac{1}{\sqrt{|\mathcal{B}|}} \sqrt{\mathbb{V}_{\mathcal{B}}[\Delta_p]} \, \eta,
    \qqquad \eta \sim \mathcal{N}(0,1),
  \end{align}
  where $\mathbb{E}_{\mathcal{B}}[\Delta_p]$ and
  $\mathbb{V}_{\mathcal{B}}[\Delta_p]$ are the mean and the variance of the
  gradient distribution of the batch respectively, and $\eta$ is Gaussian
  noise.
  Rewriting Eq.~(\ref{eq:SGD}) in terms of the mean drift and the fluctuation
  then yields
  \begin{align} \label{eq:langevin-matrix}
    W_{n+1} = W_n - \alpha \mathbb{E}_{\mathcal{B}}[\Delta_p] 
    + \frac{\alpha}{\sqrt{|\mathcal{B}|}} \sqrt{\mathbb{V}_{\mathcal{B}}[\Delta_p]} \,\eta,
  \end{align}
  i.e., a discrete Langevin equation for SGD.

  Note that the learning rate does not have a natural interpretation as a
  `physical' step size, as naively sending $\alpha \to 0$ does not give a
  correct stochastic differential equation \cite{mandt2015, li2017,
  yaida2018}.
  Instead, to obtain a continuous time limit that satisfies It\^o calculus, one
  must consider a combination of parameters that jointly acquire the dimension
  of time, which we plan to discuss in the future.

\section{Dynamics of eigenvalues, random matrix theory and the Coulomb gas}

  To follow the dynamics during learning, it is convenient to work with
  singular or eigenvalues, as some statistical properties of those are well
  known in RMT.
  RMT is usually defined for square symmetric or hermitian matrices. 
  Since weight matrices in ML are typically rectangular, of size $M\times N$,
  we consider the symmetric combination $X=W^TW$. 
  The update for $X$ follows directly from the update for $W$ given above,
  using $\delta X = W^T \delta W + \delta W^T W$.
  We denote the eigenvalues of $X$ with $x_i$ ($i=1, \ldots, N$), where we
  assume $N\leq M$ (if not, swap $W$ and $W^T$). Note that the eigenvalues
  $x_i$ are real and non-negative. 

  To obtain a discrete Langevin equation for the eigenvalues, we would have to
  write the matrix as a product of rotations and a diagonal matrix, and
  separate the dynamics of the eigenvalues from the dynamics of the angles,
  which is non-trivial in general.
  A key result from Dyson Brownian motion \cite{Dyson-4,Meh2004} is that the
  equation satisfied by the eigenvalues of $X$  can be written down directly in
  terms of the drift and fluctuations of $X$, as well as a Coulomb term, namely
  \cite{Aarts:2024wxi}
  \begin{align}
    x_i' = x_i + \alpha \tilde K_i
    + \frac{\alpha^2}{|\mathcal{B}|} \sum_{j \neq i} \frac{\tilde g_i^2}{x_i - x_j}
    + \frac{\alpha}{\sqrt{|\mathcal{B}|}} \sqrt{2} \tilde g_i \eta_i.
    \label{eq:eigen_update}
  \end{align}
  where $\tilde K_i = -\mathbb{E}_{\mathcal{B}} \left[ \Delta_p \right]_{ii}$
  and $2\tilde g_i^2 = \mathbb{V}_{\mathcal{B}} \left[  \Delta_p \right]_{ii} =
  2\mathbb{V}_{\mathcal{B}} \left[  \Delta_p \right]_{i\neq j}$.
  Quantities with a tilde are independent of learning rate and batch size at
  leading order. 
  The additional Coulomb-type interaction arises from the Jacobian determinant
  of the change of variables from matrix elements to eigenvalues, cf.\ the
  Vandermonde determinant.
  Intuitively, the interaction term originates from the fact that the
  orthogonal transformation that diagonalises $X$ does not necessarily
  diagonalise the noise matrix $\eta_{ij}$, as in the equivalent of
  Eq.~(\ref{eq:langevin-matrix}) for $X$.

  After having obtained the Langevin equation for the eigenvalues, we can
  proceed and solve the associated Fokker-Planck equation to study the
  stationary distribution.
  The Fokker-Planck equation reads (using continuous time here, we are mostly
  interested in the stationary solution)
  \begin{align}
    \partial_t P\left(\{x_i\}, t \right)
     = \sum_{i=1}^{N} \partial_{x_i} \left[ \frac{\alpha^2}{|\mathcal{B}|} \tilde g_i^2 \partial_{x_i} 
     - \alpha \tilde K_i - \frac{\alpha^2}{|\mathcal{B}|} \sum_{j \neq i} \frac{\tilde g_i^2}{x_i - x_j}
      \right] P \left( \{x_i\}, t\right).
  \end{align}
  The stationary distribution, $ \partial_t P\left(\{x_i\}, t \right)=0$, is
  solved using the Coulomb gas description \cite{Dyson-4,Meh2004}
  \begin{align}
    P_s(\left\{x_i \right\}) = \frac{1}{Z} \prod_{i<j} \left| x_i - x_j \right| 
    \exp\left[ -\sum_i \frac{1}{\alpha/|\mathcal{B}|} \frac{\tilde V_i \left(x_i\right)}{\tilde g_i^2} \right],
    \qqquad
    \tilde K_i(x_i) = - \frac{d \tilde V_i \left(x_i\right)}{d x_i} ,
    \label{eq:coulomb_gas}
  \end{align}
  where the potential $\tilde V_i(x_i)$ of the Coulomb gas is defined via the
  drift.
  If one assumes there is a unique minimum $x_i=x_i^s$ for each eigenvalue,
  such that the potential can be expanded as
  \begin{align}
    \tilde V_i(x_i) = \tilde V_i(x_i^s) +\frac{1}{2} \Omega_i\left(x_i-x_i^s\right)^2 +\ldots,
  \end{align}
  with $\Omega_i$ the curvature around the minimum, the Coulomb gas potential
  becomes a sum of Gaussians centred at $x_i=x_i^s$ with variance 
  \begin{align} \label{eq:width}
    \sigma_i^2 = \frac{\alpha}{|\mathcal{B}|}  \frac{\tilde g_i^2}{\Omega_i}
    =\frac{\alpha}{|\mathcal{B}|} \frac{\mathbb{V}_{\mathcal{B}} \left[ \Delta_p \right]_{ii}}{2\Omega_i},
  \end{align}
  The beauty of expression (\ref{eq:width}) is that the contributions to the
  fluctuations in the system are clearly separated into two factors from
  different sources, with the first part $\alpha/|\mathcal{B}|$ solely coming
  from the stochasticity of the optimiser, which leads to the linear scaling
  rule \cite{Goyal-1}, and the second part $\mathbb{V}_{\mathcal{B}} \left[
  \Delta_p \right]_{ii} / \Omega_i$ only depending on the specific profile of
  the model \cite{Aarts:2024wxi}.

\section{Applications}

  In this section, we validate the RMT description of matrix-valued SGD and the
  linear scaling rule by observing the eigenvalue distribution of a Gaussian
  Restricted Boltzmann Machine and explore a dense linear neural network within
  the teacher-student setting.

\subsection{Gaussian Restricted Boltzmann Machine}

\begin{figure}[t]
    \begin{center}
      \includegraphics[width=0.6\textwidth]{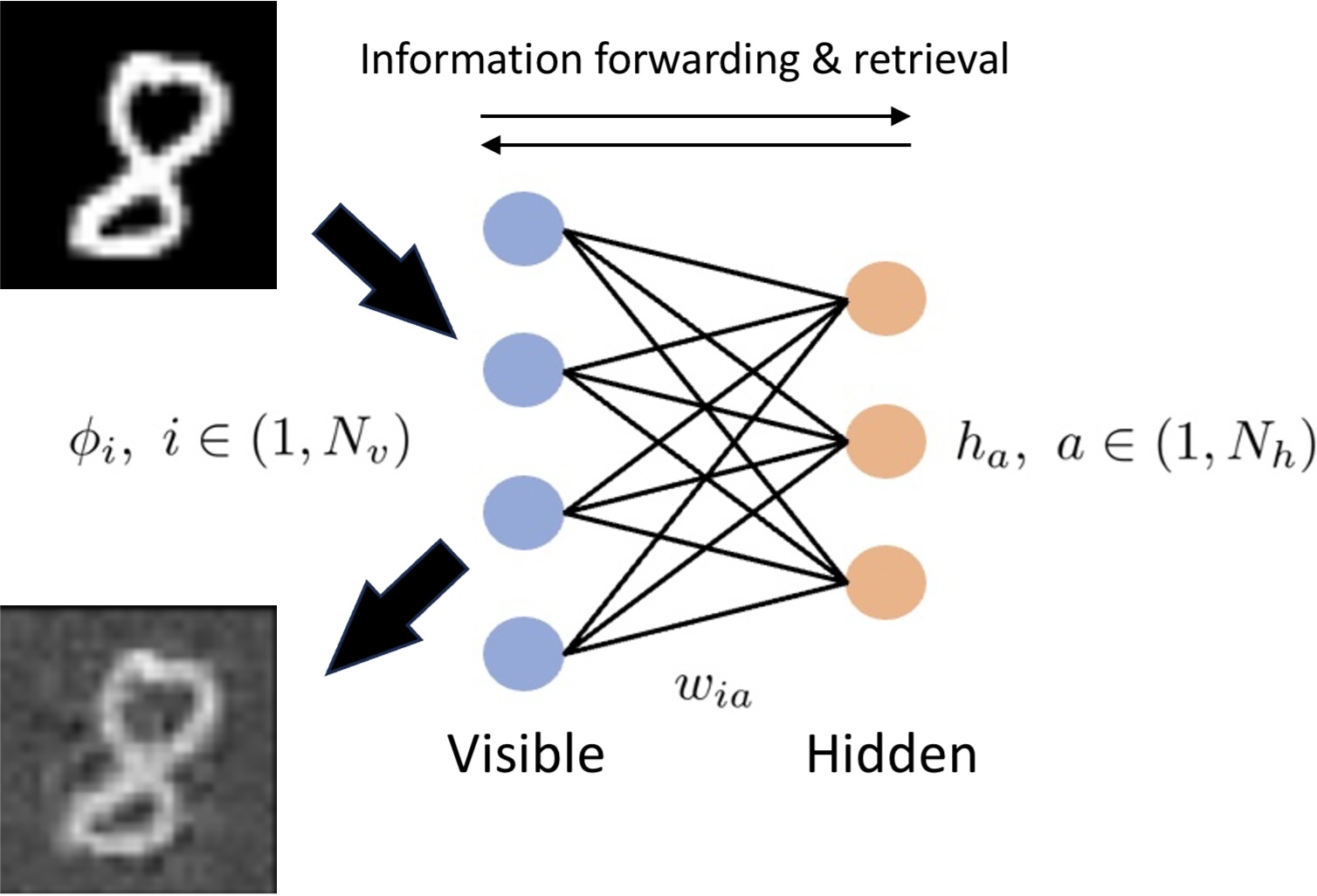}
      \caption{General structure of a Restricted Boltzmann Machine, with $N_v$ ($N_h$) visible (hidden) nodes.}
      \label{fig:rbm}
    \end{center}
  \end{figure}

  Restricted Boltzmann Machines (RBMs) are generative energy-based models
  consisting of two layers as shown in Fig.~\ref{fig:rbm} \cite{smolensky,
  10.1162/089976602760128018, Decelle_2021}. 
  When a sample is fed into the visible layer, the value of the hidden layer is
  sampled from a conditional probability distribution obtained by marginalising
  visible degrees of freedom of the given energy function of the model.
  Subsequently the output of the model is sampled from a conditional
  probability distribution obtained by marginalising the hidden degrees of
  freedom.
  The model is stochastically updated by maximising the log-likelihood between
  the model distribution and the target distribution.
  If both the visible and hidden degrees of freedom are Gaussian fields,
  interacting via a bilinear coupling, $\phi_i W_{ia}h_a$, one obtains a
  Gaussian RBM, with a probability distribution $p(\phi,h)\sim
  \exp[-S(\phi,h)]$ and the ``action'', 
  \begin{align}
    S(\phi, h) = \half \mu^2 \phi^T\phi +\frac{1}{2\sigma_h^2}(h-\eta)^T(h-\eta) - \phi^T W h.
  \end{align}
  Here $\mu^2$ and $\sigma_h^2$ are hyperparameters and we put the bias
  $\eta=0$.
  A full analysis of this model using the language of LFT can be found in
  Ref.~\cite{Aarts:2023uwt}.

  We train the Gaussian RBM to learn a distribution representing a
  one-dimensional non-interacting lattice scalar field theory, i.e., the
  eigenvalues of the target distribution are given as $\kappa_n = m^2 + 2 - 2
  \cos \left(2 \pi n/N \right)$, where $m$ is the mass of the scalar field,
  $N=N_v$ is the size of the lattice and $-N/2<n\leq N/2$. 
  The smallest and largest eigenvalues are non-degenerate, while all the
  intermediate ones are doubly degenerate.
  We denote the learnt RBM eigenvalues as $\lambda_i=\mu^2-x_i$, where
  $x_i=\sigma_h^2\xi_i^2$, with $\xi_i$ the singular values of $W$
  \cite{Aarts:2023uwt,Aarts:2024wxi}.

  \begin{figure}[ht!]
    \centering
    \includegraphics[width=0.48\linewidth]{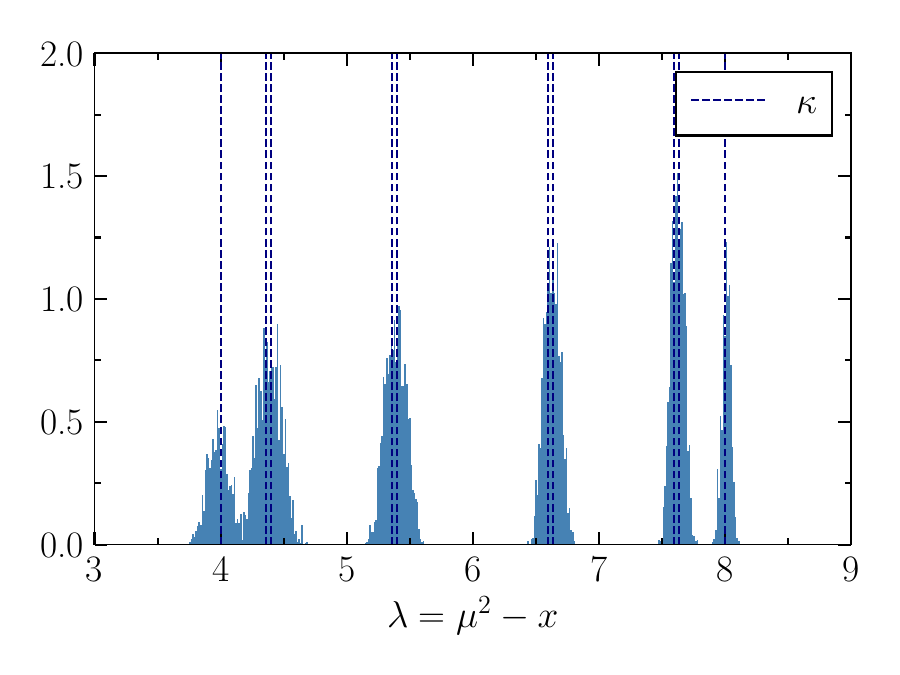}
    \includegraphics[width=0.48\linewidth]{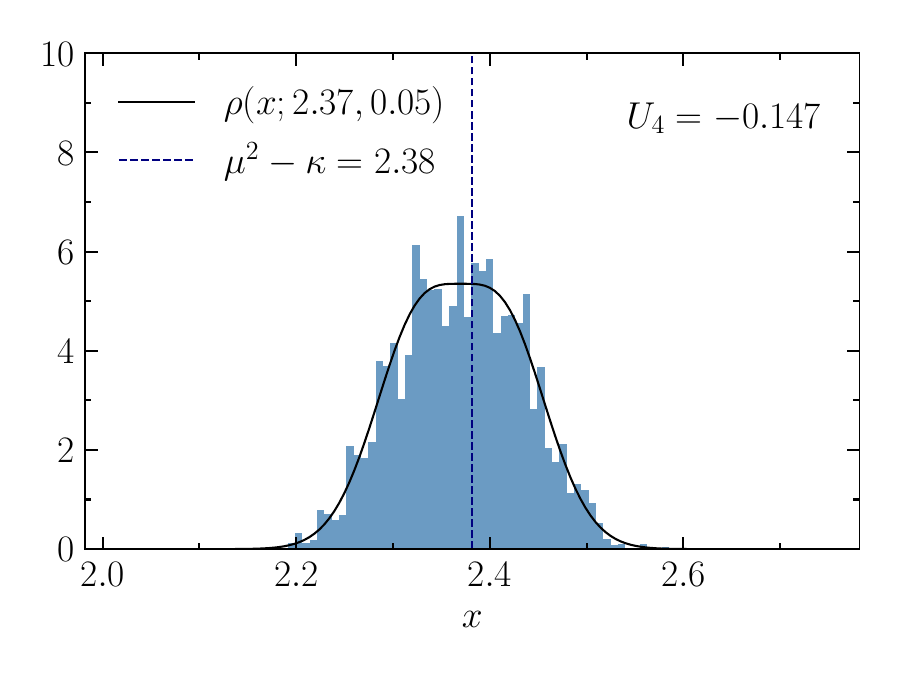}
    \caption{(Left) Target eigenvalues (dashed lines) and model eigenvalues
    (histograms) after training. The middle $8$ target eigenvalues are doubly
    degenerate due to periodic boundary conditions. (Right) Close-up of one of
    the peaks: the learnt eigenvalue distribution of the RBM follows the Wigner
    semi-circle (solid line). }
    \label{fig:hist}
    \centering
    \includegraphics[width=0.48\linewidth]{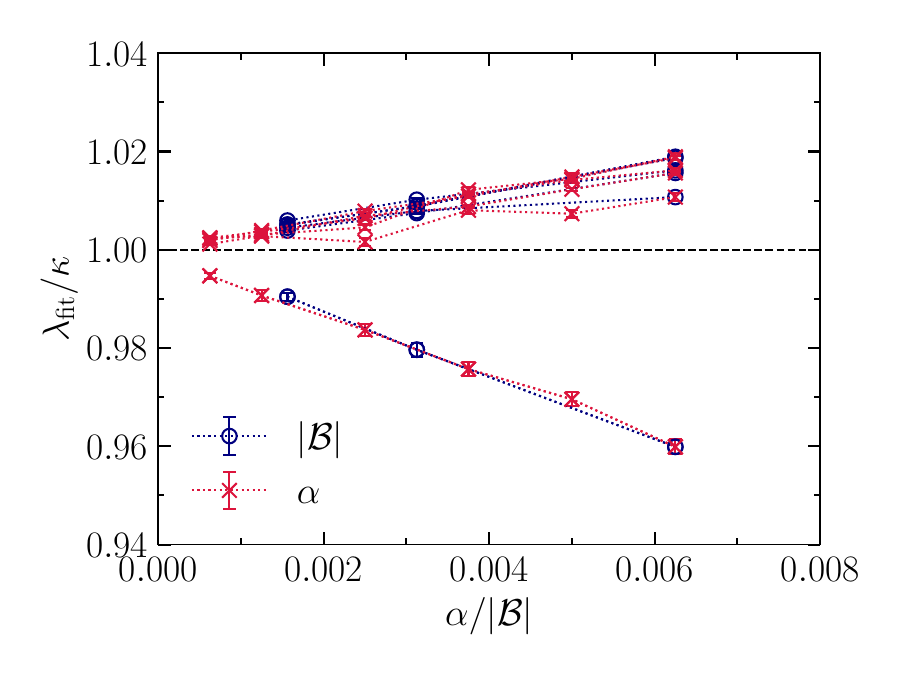}
    \includegraphics[width=0.48\linewidth]{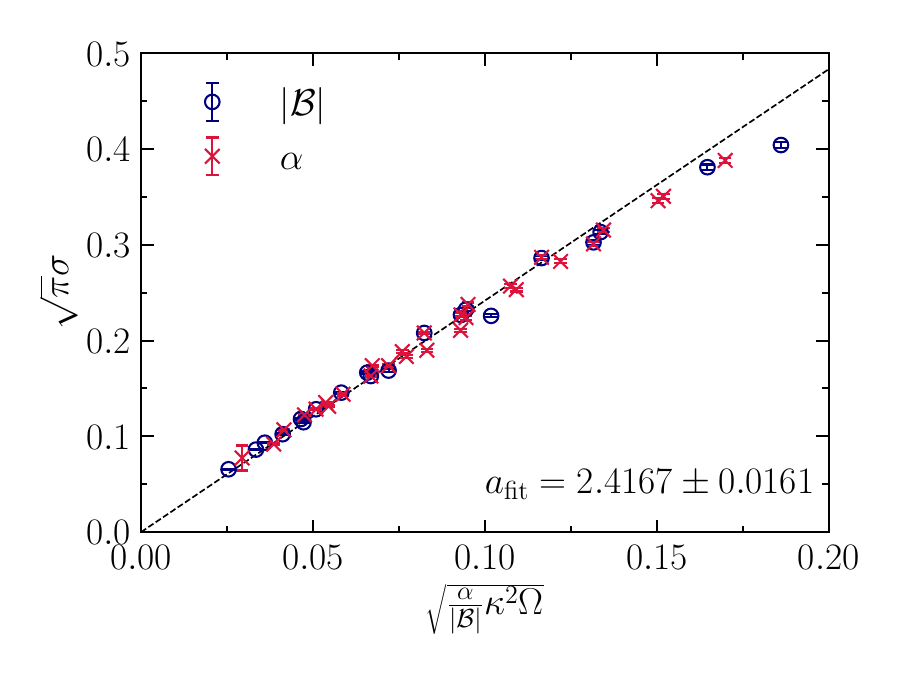}
    \caption{(Left) Deviation of the fitted centres of the model eigenvalue
    distributions from the target ones due to the inter-mode Coulomb
    interaction. The offset decreases as stochasticity in the model decreases.
    (Right) The width of eigenvalue distributions scales with the predicted
    universal scaling factor $\sqrt{\alpha/|\mathcal{B}|}$. In both cases
    $\alpha$ and $|\mathcal{B}|$ are varied independently.
    }
    \label{fig:fit}
  \end{figure}

  After training, the eigenvalues of the RBM flow towards the target
  eigenvalues, but rather than learning the exact values, they form a
  distribution around the target values, as shown in Fig.~\ref{fig:hist}
  (left). 
  The precise structure of the distributions can be analysed using RMT
  \cite{Aarts:2024wxi}. 
  In Fig.~\ref{fig:hist} (right) we show one of the distributions (note that
  $\mu^2=9$ and hence $x=9-\lambda$). Using the Coulomb gas description, the
  predicted spectral density, a.k.a.\ the Wigner semi-circle, for each pair of
  doubly-degenerate eigenvalues, is 
  \begin{align} \label{eq:spectral_density}
    \rho (x; x_m, \sigma) =  \frac{1}{N} \sum_{i=1}^{N} \left\langle\delta (x - x_i) \right\rangle 
    \stackrel{N=2}{=} 
    \frac{e^{-\delta x^2 / \left(2\sigma^2\right)}}{4\sqrt{\pi} \sigma}
    \left[
       2 e^{-\delta x^2 / \left(2 \sigma^2\right)} + \sqrt{2 \pi} \frac{\delta x}{\sigma}
       \text{Erf} \left( \frac{\delta x}{\sqrt{2} \sigma} \right)
    \right],
  \end{align}
  where $\delta x=x-x_m$, with $x_m$ the centre of the distribution, and
  $\sigma\sim \sqrt{\alpha/|\mathcal{B}|}$ is given by Eq.~(\ref{eq:width}) and
  used as a fit parameter. 
  Fig.~\ref{fig:hist} (right) shows the fit of this spectral density to one of
  the histograms. 
  To illustrate that the histogram is indeed not a Gaussian we also compute the
  Binder cumulant $U_4$,
  \begin{align}
    U_4 \equiv \frac{\left\langle \delta x^4 \right\rangle}{3\left\langle \delta x^2 \right\rangle^2} - 1 = -\frac{4}{27} \approx -0.148 \dots
    \quad \text{for the Wigner semi-circle,}
  \end{align}
  and find excellent agreement. 
  One may note in Fig.~\ref{fig:hist} (right) that the centre of the peak is
  slightly displaced from the target value. 
  This is due to the Coulomb repulsion between all the modes, as demonstrated
  in Fig.~\ref{fig:fit} (left), where the ratio of the fitted peak centres and
  the target values is shown.
  Only in the limit of vanishing stochasticity ($\alpha/|\mathcal{B}|\to 0$) is
  the spectrum learnt exactly.  

  The universal appearance of $\alpha/|\mathcal{B}|$ is also demonstrated in
  Fig.~\ref{fig:fit} (right), where the dependence of the width $\sigma$ of the
  spectral density on $\sqrt{\alpha/|\mathcal{B}|}$ is shown, including the
  non-universal (model-dependent) factor $\kappa^2\Omega$.  
  Finally, the spacing between nearest eigenvalues follows the Wigner surmise,
  $P(s) = (\pi/2) s \exp(-\pi s^2/4)$,
  where $s=S/\langle S\rangle$ and $\langle S\rangle = \sqrt{\pi} \sigma\sim
  \sqrt{\alpha/|\mathcal{B}|}$, see Ref.~\cite{Aarts:2024wxi} for details. 

\subsection{Neural network in teacher-student setting}

  To extend the analysis to more general architectures, we consider here the
  simplest case of a neural network with one hidden layer in the
  teacher-student setting, with the activation function put equal to the
  identity. 
  Even though this is a linear network, it introduces a new feature in the
  spectral density, which is already interesting to understand, since it
  indicates how the architecture may interplay with the RMT structure seen
  above.  

  Teacher-student models are widely used in the statistical mechanics of
  learning, see e.g.\
  Refs.~\cite{annurev:/content/journals/10.1146/annurev-conmatphys-031119-050745,Goldt:2020}.
  They are formulated in terms of a teacher and a student network, with weight
  matrices $W_t$ and $W_s$ respectively. 
  The input data $\mb{x}$ is drawn from a normal distribution with unit
  variance. 
  The teacher network has fixed weights and it is the task of the student to
  determine these weights. Denoting the output of the network with $\mb{y}_p =
  \mb{f}(\mb{x}_p; W)$ for each data point $\mb{x}_p$, this can be summarised
  as
  \begin{align}
    \mathcal{L}(W_s) = \frac{1}{2P} \sum_{p=1}^{P} \left| \mb{y}_p^{(t)} - \mb{y}_p^{(s)} \right|^2, 
    \qqquad
    \mb{y}_p^{(t)} = \mb{f}(\mb{x}_p; W_t), \qqquad \mb{y}_p^{(s)}= \mb{f}(\mb{x}_p; W_s), 
  \end{align}
  where the sum in the loss function is over the $P$ data points. 

  The one-hidden-layer network function we use here can be written as
  \begin{align}
    \mb{f}(\mb{x}; W) = Z \, \mb{a}(W\mb{x}).
  \end{align}
  Bold-faced quantities are vectors; $W$ and $Z$ are rectangular matrices in
  general.
  The input and output dimensions do not have to be the same. 
  As stated, we replace the activation function $\mb{a}(\cdot)$ with the
  identity and we will not write it from now on. 
  The $Z$ matrix is the same for the teacher and the student, such that only
  $W$ needs to be learnt. 

  The gradient of the loss function for a data point $\mb{x}$ with components
  $x_i$ is given by 
  \begin{align}
     \frac{\partial \mathcal{L}(W_s)}{\partial {W}_{s,i'i}} = - \sum_{j', j} (Z^TZ)_{i'j'} \left(W_t-W_s\right)_{j'j} x_j x_i,
  \end{align}
  where primed and unprimed indices may have a different range, reflecting that
  the matrices are typically rectangular. 
  Averaging over a mini-batch, we can write 
  \begin{align}
    \frac{1}{|\mathcal{B}|} \sum_{p \in \mathcal{B}} x_{i,p} x_{j,p} \simeq \delta_{ij}.
  \end{align}
  This approximation ignores some stochasticity due to mini-batch sampling, but
  if the batches are not too small, we found that this can safely be ignored.
  The gradient of the loss function, averaged over a mini-batch, then reads
  \begin{align}
    \frac{\partial \mathcal{L}(W_s)}{\partial {W}_{s,i'i}} \bigg|_{\mathcal{B}} = - \sum_{j'}(Z^TZ)_{i'j'} \left(W_t-W_s\right)_{j'i}.
  \end{align}
  To analyse the dynamics analytically, we assume that some continuous time
  limit exists, while keeping $\alpha$ as an explicit learning rate, and
  consider the equation
  \begin{align}
    W_s' = W_s - \alpha \frac{\partial \mathcal{L}(W_s)}{\partial W_s} \bigg|_{\mathcal{B}} \qquad \Rightarrow\qquad
   \dot W_s = \alpha (Z^TZ) (W_t-W_s).
   \label{eq:W_update}
  \end{align}
  Here the dot indicates the time derivative and we no longer write the indices
  explicitly. 

  We can study this dynamics in more detail by introducing a singular value
  decomposition of both the teacher and the student weight matrix, writing 
  \begin{align}
    W_s = U_s\Xi_s V^T_s, \qqquad  W_t = U_t\Xi_t V^T_t,
  \end{align}
  where $U_{s,t}$ and $V_{s,t}$ are orthogonal matrices and $\Xi_{s,t}$ are
  diagonal matrices containing the singular values $\xi_{s,i}$ and $\xi_{t,i}$. 
  Closely following Ref.~\cite{Aarts:2023uwt}, we take the time derivative of
  $W_sW_s^T$ and conjugate the resulting expression with $U_s^T$ and $U_s$.
  This yields
  \begin{align}
    & U_s^T \frac{d}{dt} (W_sW_s^T) U_s = 
   \frac{d}{dt} ( \Xi_s\Xi_s^T ) + (U_s^T \dot U_s) (\Xi_s\Xi_s^T) + (\Xi_s \Xi_s^T) (\dot U_s^T U_s) \nn \\
    & = \alpha ( U_s^T Z^TZ U_s ) \left[ U_s^T U_t \Xi_t V_t^T V_s \Xi_s^T - \Xi_s\Xi_s^T \right]
    + \alpha  \left[ \Xi_s V_s^T V_t \Xi_t^T U_t^T U_s - \Xi_s\Xi_s^T \right] ( U_s^T Z^TZ U_s ).
    \label{eq:dyn}
  \end{align}
  A similar expression is obtained starting from $W_s^TW_s$, but with $V_s^T
  \dot V_s$ instead of $U_s^T \dot U_s$, etc.
  From these expressions we can clearly see the process of learning: the RHS of
  the equation vanishes when the singular values of $W_s$ and $W_t$ agree,
  $\Xi_s\to \Xi_t$, as well as the left and right basis, $U_s\to U_t, V_s\to
  V_t$.
  The rate of learning is determined by both $\alpha$ and the combination
  $U_s^T Z^TZ U_s$. 
  Note that the final two terms on the first line form a symmetric matrix with
  zeroes on the diagonal, whereas the first term is purely diagonal. 

  We focus here on the diagonal terms, i.e.\ the eigenvalues of $\Xi_s\Xi_s^T$,
  or equivalently the square of the singular values of $\Xi_s$.
  We denote the eigenvalues of the student matrix with $x_i=\xi_{s,i}^2$, as in
  the RBM (there should be no confusion with the data points $\mb{x}$ discussed
  above). 
  Assuming that $U_s\sim U_t$, $V_s\sim V_t$, the diagonal part of
  Eq.~(\ref{eq:dyn}) can then be reduced to
  \begin{align} \label{eq:xdot}
    \dot x_i = 2a_i \left( \sqrt{\kappa_i x_i} - x_i\right),
  \end{align}
  where $a_i> 0$ depends on $\alpha$ times the diagonal components of $U_s^T
  Z^TZ U_s$, and $\sqrt{\kappa_i}$ is the $i^{\rm th}$ singular value of the
  teacher matrix $\Xi_t$.
  This equation is solved as
  \begin{align}
    x_i(t)  = \left[  \sqrt{\kappa_i} + \left( \sqrt{x_{i,0}} -\sqrt{\kappa_i} \right) e^{-a_i t}  \right]^2,
    \label{eq:fit}
  \end{align}
  with $x_i(t\to\infty)=\kappa_i$, as expected for learning in the absence of
  stochasticity. 
  Initialising $W_s$ from a normal distribution with variance 1, the initial
  average value of $x_i$ equals $\langle x_{i,0}\rangle =1$.
  Note that the dynamics is linear for the singular values, 
  \begin{align}
    \dot \xi_{s,i} = a_i \left( \sqrt{\kappa_i} - \xi_{s,i}\right),
    \qqquad
    \xi_{s,i}(t)  = \sqrt{\kappa_i} +\left( \xi_{s,i}(0) - \sqrt{\kappa_i} \right) e^{-a_i t}.
  \end{align}
  For reference to the Coulomb gas description, we note that the drift, i.e.,
  the RHS of Eq.~(\ref{eq:xdot}) can be obtained from a potential,
  \begin{align}
    V(x_i) = a_i\left(x_i^2-\frac{4}{3}\sqrt{\kappa_i} x_i^{3/2}\right).
    \label{eq:V}
  \end{align}
  Expanding this potential around the minimum at $x_i=\kappa_i$ yields
  \begin{align}
    V(x_i) = -\frac{1}{3} a_i\kappa_i^2 + \frac{a_i}{2} \left( x_i-\kappa_i\right)^2 + \ldots,
  \end{align}
  i.e., the width of the potential is proportional to the rate of learning $a_i$.

  \begin{figure}[t]
    \centering
    \includegraphics[width=0.48\linewidth]{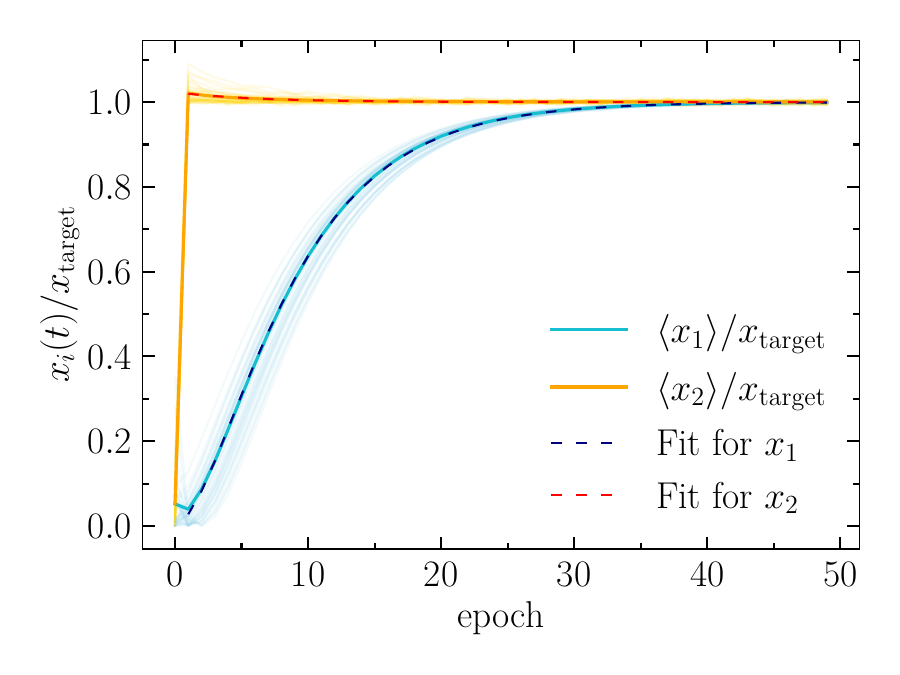}
    \includegraphics[width=0.48\linewidth]{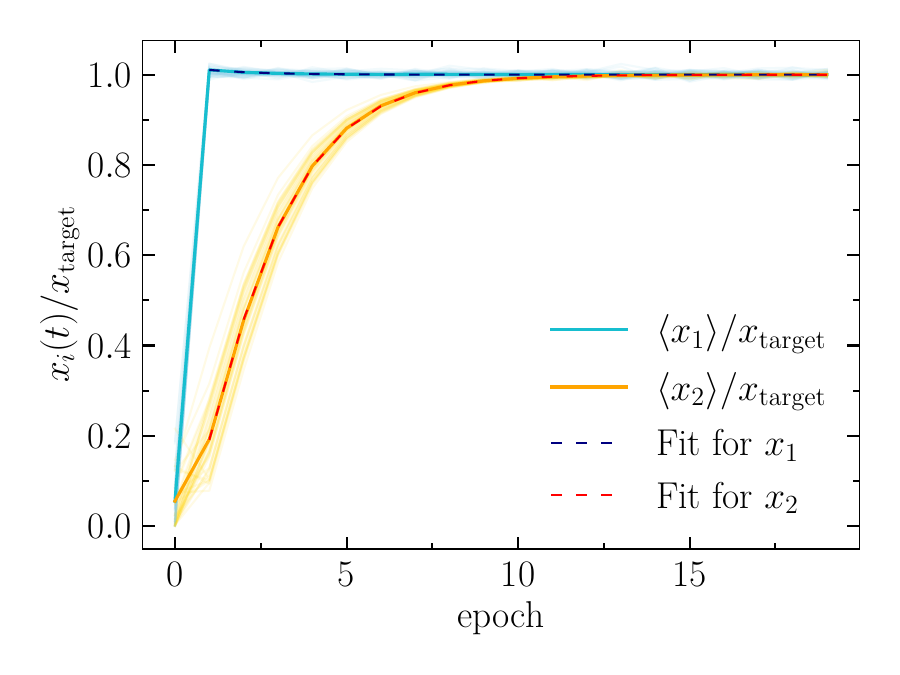}
    \caption{ Training dynamics of the square of the singular values of a $2
    \times 2$ student matrix, given a teacher matrix with doubly degenerate
    eigenvalues, using $Z$  as in Eq.~\eqref{eq:mat_1} (left) and
    Eq.~\eqref{eq:mat_2} (right). The presence of $Z$ affects the rate of
    convergence. Shown are an ensemble of 20 networks (with opaque lines), the
    evolution averaged over an ensemble of 500 networks (with solid blue and
    orange lines), and fits to Eq.~\eqref{eq:fit}, starting from epoch $t=2$
    (with dashed lines), agreeing with the averaged evolution. }
    \label{fig:eigen}
  \end{figure}

  We solved the teacher-student model numerically using SGD for the case of
  two-dimensional input and output, and with $Z$ and $W$ $2\times 2$ matrices. 
  To study the case of (near-)degeneracy, the eigenvalues of $W_t$ are taken
  identical.
  The modes mix due to the presence of $Z$.
  We study two examples, with
  \begin{equation}
    Z =
    \begin{pmatrix}
    -0.1727 & -0.8341 \\
    -0.0106 & -0.7750
    \end{pmatrix}, 
    \qqquad 
    Z^T Z =
    \begin{pmatrix}
    0.0299 & 0.1522 \\
    0.1522 & 1.2964
    \end{pmatrix},
    \label{eq:mat_1}
  \end{equation}
  and with 
  \begin{equation}
    Z =
    \begin{pmatrix}
    -1.8198 & 0.5861 \\
    -0.6574 &  0.0473
    \end{pmatrix}, 
    \qqquad
    Z^T Z =
    \begin{pmatrix}
    3.7437 & -1.0977 \\
    -1.0977 &  0.3458
    \end{pmatrix}.
    \label{eq:mat_2}
  \end{equation}
  We add a noise term in the learning process, i.e., the update rule for $W$ is
  modified as
  \begin{equation}
    W'_s = W_s + \alpha (Z^TZ) (W_t-W_s - \eta), 
    \qqquad
    \eta \sim \mathcal{N}(0, 0.01).
  \end{equation}
  This is required, as the algorithm by itself is not noisy enough, unlike in
  the RBM discussed above.
  The respective eigenvalue dynamics is shown in Fig.~\ref{fig:eigen}. 
  We used here a finite learning rate and batch size, and associate epoch with
  time. 
  The eigenvalue converging faster is associated with the larger value on the
  diagonal of $Z^T Z$ and the fit obtained using Eq.~\eqref{eq:fit} aligns with
  the ensemble average of the eigenvalue dynamics. 
  When $Z= 1\!\!1$, the rate of convergence is the same for all eigenvalues.

  To analyse these results in terms of RMT and the Coulomb potential, we note
  that  the additional layer in the network leads to different rates of
  learning and curvatures in the potential (\ref{eq:V}), parametrised by $a_i$.
  We incorporate this by extending the RMT description to a two-component
  Coulomb gas, where particles of each species are characterised by a different
  mass (or variance). 
  We focus on the case of doubly degenerate eigenvalues, with $\kappa_1 =
  \kappa_2 = \kappa$.
  By shifting $x_{1,2}$ by $\kappa$ and extending the integration boundaries to
  $-\infty$, we arrive at the Coulomb gas partition function
  \begin{align}
    Z = \frac{1}{N_0} \int dx_1 dx_2\, |x_1-x_2| e^{-V(x_1,x_2)}, 
    \qqquad
    V(x_1, x_2) = \frac{x_1^2}{2\sigma_1^2} + \frac{x_2^2}{2\sigma_2^2},
  \end{align}
  where the two modes have different variances $\sigma_{1,2}^2$.
  The normalisation constant is $N_0 = 4\sqrt{\pi}\sigma_1\sigma_2\sigma_m$,
  with
  \begin{align}
    \sigma_m^2 = \half\left(\sigma_1^2+\sigma_2^2\right). 
    \label{eq:sigma_relation}
  \end{align}
  We follow the same procedure as in the RBM. 
  For the Wigner surmise, we write  $S=x_1-x_2$, $x_c = \sigma_2/(2\sigma_1)
  x_1 + \sigma_1/(2\sigma_2)x_2$, such that $V(x_1, x_2)  = S^2/(4\sigma_m^2) +
  x_c^2/\sigma_m^2$.
  The Wigner surmise and the average level spacing are then 
  \begin{align} \label{eq:S_distribution}
    P(S) = \frac{S}{2\sigma_m^2} \exp[-S^2/(4\sigma_m^2)],
    \qqquad
    \langle S \rangle = \sqrt{\pi}\sigma_m.
  \end{align}
  The spectral density is given by
  \begin{align}
    \rho(x; x_c, \sigma_1, \sigma_2) =
    \frac{e^{-\sigma_m^2 \delta x^2/(\sigma_1^2\sigma_2^2)}}{8\sqrt{\pi}\sigma_1 \sigma_2\sigma_m  } 
    \sum_{i=1, 2}
    \left[ 
    2\sigma_i^2 + e^{\delta x^2/(2\sigma_i^2)} \sqrt{2\pi}\delta x\sigma_i \text{Erf}\left(\frac{\delta x}{\sqrt{2}\sigma_i}\right)\right],
    \label{eq:gen_wigsc}
  \end{align}
  where $\delta x=x-x_c$ and $\sigma_m$ is related to $\sigma_{1,2}$ using
  Eq.~(\ref{eq:sigma_relation}). 
  The presence of the additional layer affects the spectral density, which
  becomes a generalised version of the Wigner semi-circle, whereas the Wigner
  surmise is unchanged, albeit with a modified factor for the average level
  spacing.

  \begin{figure}[t]
    \centering
    \includegraphics[width=0.48\linewidth]{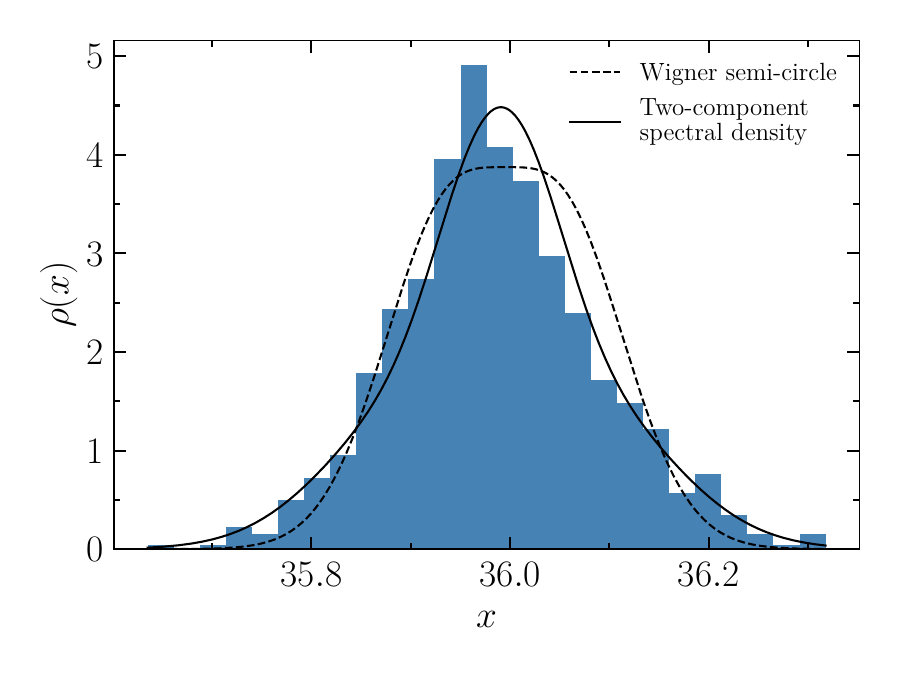}
    \includegraphics[width=0.48\linewidth]{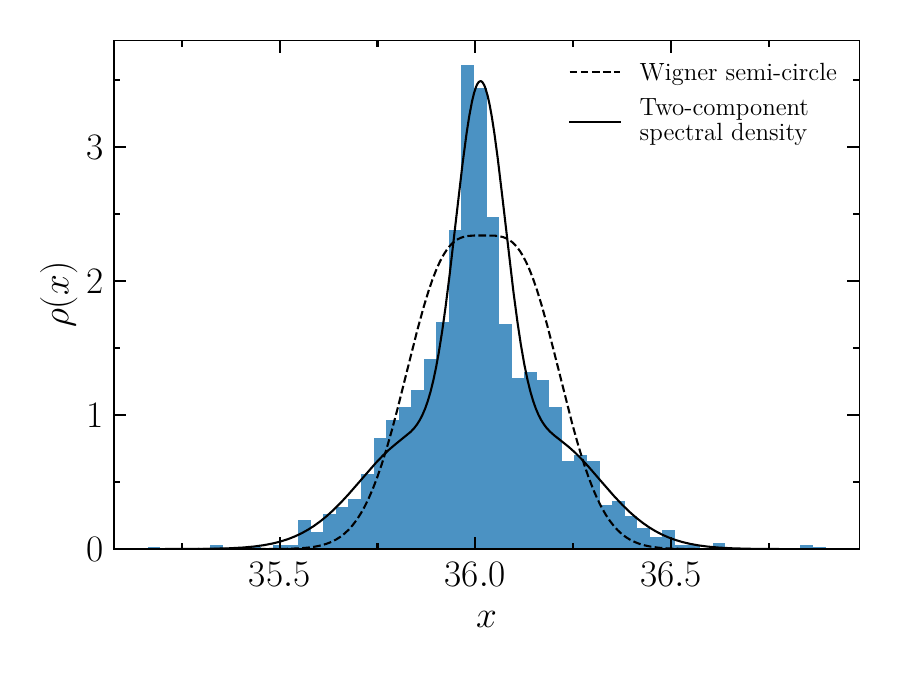}
    \caption{ Histogram of the spectral density $\rho(x)$ in the presence of a
    hidden layer, with $Z$ as in Eq.~\eqref{eq:mat_1} (left) and
    Eq.~\eqref{eq:mat_2} (right). Also shown are fits to the standard Wigner
    semi-circle \eqref{eq:spectral_density} (dashed line) and the generalised
    Wigner semi-circle \eqref{eq:gen_wigsc} for two species (solid line). The
    generalised Wigner semi-circle better captures the histogram's peak and
    wider tails,  as seen in particular on the right. }
    \label{fig:spectral_densities}
  \end{figure}

  We have verified this Coulomb gas description for both the Wigner surmise
  (not shown here) and the generalised Wigner semi-circle. 
  In Fig.~\ref{fig:spectral_densities}, we show the histogram of the
  eigenvalues and a comparison between the fits performed with the standard
  Wigner semi-circle~\eqref{eq:spectral_density} and its generalised
  version~\eqref{eq:gen_wigsc}. It is evident that the standard expression is
  not able to reproduce the high peak and the wider tails, but that the
  generalised version is. 
  Finally we note that inserting the values of $\sigma_{1,2}$ into
  Eq.~\eqref{eq:sigma_relation}, we obtain a result for $\sigma_m$ compatible
  with the value obtained directly from a fit of the surmise to
  Eq.~\eqref{eq:S_distribution}.

\section{Summary}

  To further develop our understanding of machine learning algorithms, we have
  formulated stochastic gradient descent in terms of Dyson Brownian motion and
  the Coulomb gas.  
  In the stationary limit the statistical properties of singular/eigenvalues of
  weight matrices then follow predictions from random matrix theory. 
  In particular,  we have shown that the width of the Coulomb potential around
  a learnt target value scales proportionally to a specific combination of two
  hyperparameters of the optimiser, namely the learning rate over batch size,
  and hence derived the linear scaling rule.

  We have verified this behaviour in the Gaussian Restricted Boltzmann Machine,
  in which the spectral density takes the form of the Wigner semi-circle and
  the level spacing follows the Wigner surmise, and the predicted scaling of
  the eigenvalue distribution and level spacing with the learning rate and
  batch size is observed. 
  Subsequently we have extended the analysis into a more general scenario by
  considering a linear neural network with one hidden layer in a
  teacher-student setting. 
  Interestingly, the additional layer modifies the width of the potential for
  each eigenvalue, resulting in a Coulomb gas with multiple species and a
  generalised Wigner semi-circle.

  For the future we plan to consider larger non-linear neural networks, in
  which the spectral density is expected to be more intricate, as seen in e.g.\
  Refs.~\cite{Martin-2019, Baskerville,Aarts:2024qey}.

  \vspace*{0.3cm} 

\noindent
{\bf Acknowledgements} --  
GA, MF and BL are supported by STFC Consolidated Grant ST/T000813/1. 
BL is further supported by the UKRI EPSRC ExCALIBUR ExaTEPP project EP/X017168/1.
CP is supported by the UKRI AIMLAC CDT EP/S023992/1.

\noindent
{\bf Research Data and Code Access} --
The code and data used for in the first part of this manuscript are available from Ref.~\cite{park_2024_13310439}.

\noindent
{\bf Open Access Statement} -- For the purpose of open access, the authors have applied a Creative Commons Attribution (CC BY) licence to any Author Accepted Manuscript version arising.

%\bibliographystyle{JHEP}
%\bibliography{paper}

\providecommand{\href}[2]{#2}\begingroup\raggedright\endgroup

\end{document}